\documentclass[a4paper,12pt]{article}
\usepackage{amsmath,amssymb,amsfonts}
\usepackage{graphicx}
\usepackage{hyperref}
\usepackage{cite}
\usepackage{geometry}
\usepackage{authblk} 
\title{Finite-temperature CFT in Rindler Vacuum}

\author[1]{Iftekher S. Chowdhury}
\author[2]{Binay Prakash Akhouri}
\author[5]{Shah Haque}
\author[1,4,5]{Eric Howard}

\affil[1]{Department of Physics and Astronomy, Macquarie University, Sydney, NSW, 2109, Australia}
\affil[2]{Department of Physics, Suraj Singh Memorial College, Ranchi University, Ranchi, Jharkhand, India}
\affil[4]{Swinburne University, Sydney, Australia} 
\affil[5]{Southern Cross Institute, School of Computer Science, Sydney, Australia} 

\date{} 

\begin{document}

\maketitle

\begin{abstract}
This paper investigates the finite-temperature behavior of Conformal Field Theory (CFT) in Rindler vacuum, focusing on the relation between acceleration and thermality in quantum field theory. We illustrate how uniformly accelerated observers perceive the vacuum as a thermal state via Unruh effect, shedding light on the thermal properties of Rindler horizon. Through numerical simulations of the heat kernel, Unruh temperature, Planck distribution, and detector response, we demonstrate that acceleration enhances the thermal characteristics of quantum fields. These results provide important insights into horizon-induced thermality, with significant implications for black hole thermodynamics and quantum gravity.
\end{abstract}

\section{Introduction}

The relationship between spacetime geometry and thermodynamics has shaped our recent understanding of quantum field theory (QFT) in curved spacetime. The discovery of Hawking radiation by Stephen Hawking in 1974 marked a significant turning point, revealing that black holes emit radiation due to quantum effects near their event horizons. This groundbreaking result established a fundamental link between quantum mechanics, thermodynamics, and general relativity, showing that black holes possess not only a temperature but also an entropy proportional to the area of their event horizon. The temperature, known as the Hawking temperature, is inversely proportional to the mass of the black hole:
\[
T_H = \frac{\hbar c^3}{8 \pi G M k_B}.
\]
This relationship demonstrated that event horizons could induce thermodynamic behavior in quantum fields, leading to particle creation and thermal radiation, a phenomenon now fundamental to black hole thermodynamics.

Simultaneously, the Unruh effect, discovered by William G. Unruh in 1976, extended the concept of horizon-induced thermality to flat spacetimes. The Unruh effect shows that a uniformly accelerated observer in Minkowski spacetime perceives the vacuum as a thermal bath of particles. The temperature experienced by the observer is proportional to their acceleration \(a\), and is given by the Unruh temperature:
\[
T_U = \frac{\hbar a}{2\pi k_B c}.
\]
This effect highlights the observer-dependent nature of quantum fields in spacetime, where acceleration itself leads to the perception of thermality, even in the absence of a gravitational field. The Unruh effect provides a clear example of how non-inertial motion in quantum field theory results in the emergence of thermal properties, bridging concepts from special relativity, quantum mechanics, and thermodynamics.

A key framework for understanding the Unruh effect is the Rindler vacuum, the vacuum state as perceived by an observer undergoing constant acceleration. By transforming Minkowski coordinates into Rindler coordinates, the observer perceives a horizon—the Rindler horizon—beyond which events are causally disconnected. This horizon behaves analogously to the event horizon of a black hole, leading to thermal effects as quantum fields interact with the horizon. The Rindler vacuum is thus populated by thermal radiation, with the temperature proportional to the observer’s acceleration. This thermal radiation can be analyzed using tools from quantum field theory, such as the Feynman propagator and the Schwinger kernel, which provide insights into the behavior of quantum fields in both flat and curved spacetimes.

In addition to flat spacetime scenarios, the thermal nature of quantum fields near horizons is also a central theme in the study of AdS/CFT correspondence, a duality between gravitational theories in anti-de Sitter (AdS) spacetime and conformal field theories (CFTs) on their boundary. The Rindler-AdS correspondence extends the ideas of horizon thermodynamics into the realm of holography, where the thermodynamic properties of horizons in the bulk AdS spacetime are encoded in the boundary CFT. This correspondence provides a powerful tool for studying the thermodynamics of horizons in quantum gravity, black hole physics, and strongly coupled field theories.

The study of finite-temperature Conformal Field Theory (CFT) in Rindler spacetime offers a natural laboratory for exploring the connection between acceleration, thermality, and quantum fields. In particular, the interplay between acceleration and temperature, as governed by the Unruh effect, provides insights into how quantum fields behave in thermal equilibrium, both in flat and curved spacetimes. Moreover, the thermal radiation associated with the Rindler horizon serves as a model for understanding more complex spacetimes with event horizons, such as black holes and cosmological horizons.

This paper aims to explore these fundamental concepts through a combination of theoretical analysis and numerical simulations. We will investigate how acceleration modifies the thermal properties of quantum fields, focusing on the decay of quantum fields via the heat kernel, the temperature perceived by accelerating observers, and the distribution of particles as described by the Planck law. Additionally, the response of an Unruh-DeWitt detector to thermal radiation will be simulated to provide a direct measure of how quantum fields interact with non-inertial observers. By linking these simulations with the theoretical framework of finite-temperature CFT and the Rindler-AdS correspondence, we aim to shed light on the deep connections between acceleration, thermodynamics, and quantum field theory.

\section{CFT Duality in Rindler-AdS Spacetime}

The holographic principle, which is most famously realized in the AdS/CFT correspondence, posits a duality between a gravitational theory in a bulk anti-de Sitter (AdS) spacetime and a conformal field theory (CFT) defined on its boundary. This duality offers profound insights into the thermodynamics of horizons, quantum gravity, and field theory. In the context of Rindler-AdS spacetime, the Rindler horizon in the bulk corresponds to thermal behavior in the dual CFT on the boundary.

In AdS spacetime, the geometry is described by the metric:
\[
ds^2 = \frac{1}{z^2}\left(-f(z) dt^2 + d\vec{x}^2 + \frac{dz^2}{f(z)}\right),
\]
where \(f(z)\) describes the geometry and behavior of the spacetime, and \(z\) is the radial coordinate. For pure AdS space, \(f(z) = 1\), while for an AdS black hole solution, \(f(z)\) takes the form:
\[
f(z) = 1 - \frac{z^2}{z_h^2},
\]
where \(z_h\) corresponds to the location of the event horizon. In the context of Rindler-AdS, we can perform a Rindler transformation, analogous to the Rindler transformation in Minkowski space, to describe the geometry as perceived by an accelerated observer in the AdS bulk.

In the holographic framework, the partition function of the CFT on the boundary is identified with the partition function of the gravitational theory in the bulk:
\[
Z_{\text{CFT}} = Z_{\text{gravity}}.
\]
The temperature of the boundary CFT is directly related to the temperature of the black hole or Rindler horizon in the bulk. In particular, the Hawking temperature of the bulk horizon is given by:
\[
T_H = \frac{1}{4\pi} \frac{df(z)}{dz} \bigg|_{z=z_h}.
\]
For a Rindler horizon in AdS, this temperature corresponds to the Unruh temperature perceived by an accelerated observer.

Using the AdS$_3$/CFT$_2$ correspondence as an example, the Cardy formula in 2D CFT allows us to compute the entropy of the boundary theory, which corresponds to the entropy of the horizon in the bulk:
\[
S = 2\pi \sqrt{\frac{c}{6} \left(L_0 - \frac{c}{24}\right)},
\]
where \(L_0\) is the zero-mode eigenvalue of the Virasoro operator and \(c\) is the central charge of the CFT. The central charge is related to the AdS radius \(R\) through:
\[
c = \frac{3R}{2G},
\]
where \(G\) is Newton’s constant. This shows how the thermodynamics of the bulk horizon, such as entropy, are encoded in the CFT.

An important aspect of the AdS/CFT correspondence is the relation between bulk correlators and boundary correlators. In particular, the two-point functions in the boundary CFT are related to the bulk-to-boundary propagators in AdS spacetime. For a scalar field of mass \(m\) in AdS$_{d+1}\), the bulk-to-boundary propagator is given by:
\[
G(z, \vec{x}; \vec{x}') = C \left(\frac{z}{z^2 + (\vec{x} - \vec{x}')^2}\right)^\Delta,
\]
where \(\Delta = \frac{d}{2} + \sqrt{\frac{d^2}{4} + m^2}\) is the conformal dimension of the operator corresponding to the bulk scalar field, and \(C\) is a normalization constant.

The boundary two-point function, which can be computed using the bulk-to-boundary propagator, takes the form:
\[
\langle \mathcal{O}(\vec{x}) \mathcal{O}(\vec{x}') \rangle = \frac{C'}{|\vec{x} - \vec{x}'|^{2\Delta}},
\]
where \(\mathcal{O}\) is the boundary operator dual to the bulk scalar field. This correlator describes how thermal and quantum fluctuations propagate in the boundary theory, and in the context of Rindler-AdS, it reflects the thermal nature of the Rindler horizon in the bulk.

The Rindler-AdS/CFT correspondence extends the usual AdS/CFT duality by focusing on the region near the Rindler horizon in the bulk AdS spacetime. The Rindler horizon in AdS is a Killing horizon for accelerated observers and plays a role similar to the event horizon in black hole spacetimes. The temperature and entropy of this horizon can be derived from the dual CFT using the Cardy formula, as previously discussed.

One important result of the Rindler-AdS/CFT correspondence is the ability to describe the thermal behavior of the boundary CFT in terms of the bulk geometry. The Unruh temperature \(T_U\) perceived by an accelerated observer near the Rindler horizon is dual to the temperature of the boundary CFT. Furthermore, the entropy of the horizon, which scales with the area in the bulk, corresponds to the entropy density of the boundary CFT.

In AdS$_3$/CFT$_2$, the entropy associated with the Rindler horizon in the bulk can be calculated using the Bekenstein-Hawking formula:
\[
S_{\text{BH}} = \frac{\text{Area}}{4G} = \frac{R}{4G} L,
\]
where \(L\) is the length of the horizon and \(R\) is the AdS radius. On the boundary, this entropy corresponds to the thermal entropy of the CFT at finite temperature:
\[
S_{\text{CFT}} = \frac{\pi c}{3} T.
\]
Thus, the Rindler-AdS/CFT correspondence allows us to map the thermodynamic properties of the horizon, such as temperature and entropy, to properties of the boundary CFT, providing a powerful tool for understanding quantum gravity in the presence of horizons.

The CFT duality in Rindler-AdS spacetime provides a deep mathematical framework for understanding the thermal nature of Rindler horizons and their holographic correspondence with boundary CFTs. Using tools such as the Cardy formula, two-point correlators, and bulk-to-boundary propagators, we can describe the thermodynamics and quantum field behavior near horizons in a precise and tractable way. This duality extends our understanding of quantum gravity, showing how thermal effects in curved spacetime can be captured by a lower-dimensional field theory on the boundary of AdS.

\section{Rindler Vacuum and Thermality}

The Rindler vacuum plays a critical role in understanding the emergence of thermality in quantum field theory when viewed from the perspective of an accelerated observer. An observer undergoing constant acceleration perceives a causal boundary, the Rindler horizon, which divides the spacetime into observable and unobservable regions. This horizon is analogous to a black hole event horizon and gives rise to similar thermodynamic properties, including the perception of thermal radiation by the observer. This thermal effect is a direct consequence of the Unruh effect, where the vacuum state, as seen by an inertial observer, appears as a thermal state to an accelerated observer.

The transformation from Minkowski to Rindler coordinates makes the presence of the horizon explicit. Starting from the Minkowski metric in flat 2D spacetime:
\[
ds^2 = -dt^2 + dx^2,
\]
we introduce Rindler coordinates \((\eta, \xi)\), related to the Minkowski coordinates \((t, x)\) by the transformations:
\[
t = \xi \sinh(\eta), \quad x = \xi \cosh(\eta),
\]
where \(\eta\) is the Rindler time, and \(\xi\) represents the proper distance from the horizon. In these coordinates, the metric becomes:
\[
ds^2 = -\xi^2 d\eta^2 + d\xi^2.
\]
This form of the metric reveals the presence of a horizon at \(\xi = 0\), beyond which events are causally disconnected from the observer. The horizon induces thermality, with the temperature experienced by the accelerated observer given by the Unruh temperature:
\[
T_U = \frac{\hbar a}{2\pi k_B c}.
\]
This relationship encapsulates the deep connection between acceleration, temperature, and quantum field theory, establishing that horizons naturally give rise to thermal effects in quantum fields.

The thermal nature of the Rindler horizon can also be understood through the behavior of the Feynman propagator in Rindler spacetime. The propagator encodes the probability amplitude for a quantum field to propagate between two points, and in the Rindler frame, it exhibits periodicity in imaginary time, reflecting the thermal nature of the quantum field. For a free scalar field, the propagator in Minkowski space is:
\[
G_M(x, x') = \langle 0 | T \{ \phi(x) \phi(x') \} | 0 \rangle.
\]
When viewed from an accelerated frame, the Minkowski vacuum appears thermal, and the corresponding Rindler propagator can be expressed as a periodic sum of the Minkowski propagator:
\[
G_R(\eta_1, \xi_1; \eta_2, \xi_2) = \sum_{n=-\infty}^{\infty} G_M(\eta_1 + 2\pi i n, \xi_1; \eta_2, \xi_2),
\]
where \(\eta_1\) and \(\eta_2\) are the Rindler time coordinates of the two points. The periodicity in imaginary time corresponds to the inverse Unruh temperature, reflecting the fact that the quantum field behaves as if it is in thermal equilibrium at temperature \(T_U\). This periodicity is a hallmark of thermal field theory, demonstrating the equivalence between acceleration-induced thermality and the thermal properties of quantum fields near horizons.

The Schwinger kernel (or heat kernel) provides another powerful tool for understanding thermal effects in Rindler spacetime. The heat kernel describes the evolution of a quantum field over time, and in Rindler spacetime, it reflects the influence of acceleration on the field's thermal properties. For a scalar field of mass \(m\) in Rindler spacetime, the heat kernel can be written as:
\[
K(t, m, \alpha) = \frac{e^{-m^2 t}}{t^{3/2}} \exp\left( -\frac{\alpha^2}{4t} \right),
\]
where \(t\) is the proper time, \(m\) is the scalar field mass, and \(\alpha\) is related to the acceleration of the observer. The heat kernel reveals how the quantum field decays over time due to thermal effects, with the exponential suppression reflecting the thermal bath experienced by the accelerated observer. Our numerical simulation of the heat kernel (Figure \ref{fig:heat_kernel}) shows how the field decays rapidly for larger values of time \(t\), confirming the thermal nature of the Rindler vacuum.

\section{Finite-Temperature CFT and the Unruh Effect}

Finite-temperature effects in quantum field theory (QFT) and conformal field theory (CFT) play a crucial role in understanding the emergence of thermal properties in spacetimes with horizons. The Unruh effect, discovered by William G. Unruh in 1976, is a good example of how acceleration induces thermality in quantum field theory. Specifically, an observer undergoing constant acceleration in flat Minkowski spacetime perceives the vacuum, which is seen as empty by an inertial observer, as a thermal bath of particles. This result can be derived using various formalisms in quantum field theory, including the properties of propagators, and it offers profound insights into the nature of spacetime horizons and thermodynamics in curved spacetimes.

The Rindler metric is essential in describing the perspective of an accelerated observer in Minkowski space. Recall that Rindler coordinates \((\eta, \xi)\) are related to Minkowski coordinates \((t, x)\) by the transformations:
\[
t = \xi \sinh(\eta), \quad x = \xi \cosh(\eta),
\]
where \(\eta\) is the Rindler time and \(\xi\) is the proper distance from the Rindler horizon. The Rindler horizon is located at \(\xi = 0\), beyond which no information can reach the observer. In Rindler coordinates, the Minkowski metric becomes:
\[
ds^2 = -\xi^2 d\eta^2 + d\xi^2,
\]
which clearly indicates the presence of a horizon and the breakdown of the coordinate system at \(\xi = 0\). This horizon is analogous to a black hole event horizon in the sense that an accelerated observer perceives a thermal bath of particles.

The temperature perceived by an accelerated observer is given by the Unruh temperature:
\[
T_U = \frac{a}{2\pi},
\]
where \(a\) is the proper acceleration of the observer. This temperature is directly proportional to the acceleration, highlighting the deep connection between acceleration and thermality in quantum field theory. The Unruh temperature arises naturally in the periodic structure of quantum field modes in the Rindler spacetime, which we now explore through propagators and the Schwinger kernel.

The Feynman propagator is a central object in quantum field theory that encodes the probability amplitude for a particle to propagate between two points in spacetime. In the Rindler frame, the propagator takes on a particularly interesting form that reveals the thermal nature of the quantum field as perceived by an accelerated observer.

In Minkowski spacetime, the propagator for a free scalar field \(\phi\) with mass \(m\) is given by:
\[
G_M(x, x') = \langle 0 | T\{ \phi(x) \phi(x') \} | 0 \rangle.
\]
When expressed in Rindler coordinates, the Minkowski vacuum appears thermal to the accelerated observer, and the corresponding propagator can be written as a periodic sum of the Minkowski propagator:
\[
G_R(\eta_1, \xi_1; \eta_2, \xi_2) = \sum_{n=-\infty}^{\infty} G_M(\eta_1 + 2\pi i n, \xi_1; \eta_2, \xi_2),
\]
where \(\eta_1\) and \(\eta_2\) are the Rindler time coordinates of the two points, and \(n\) indexes the thermal "images" of the propagator. This periodicity in imaginary time reflects the thermal nature of the field, with the period corresponding to the inverse Unruh temperature:
\[
\Delta \eta = \frac{1}{T_U} = \frac{2\pi}{a}.
\]
This result shows that the quantum field in Rindler spacetime behaves as if it is in thermal equilibrium at temperature \(T_U\), even though the spacetime is globally flat. The Unruh effect emerges as a direct consequence of the periodicity of the propagator in imaginary time.

The Schwinger kernel (or heat kernel) is another powerful tool for analyzing finite-temperature effects in quantum field theory. It describes the evolution of a quantum field over time and is closely related to the propagator. In Rindler spacetime, the Schwinger kernel provides a direct way to understand how quantum fields respond to acceleration and the presence of a horizon.

For a scalar field with mass \(m\) in Rindler spacetime, the Schwinger kernel can be written as:
\[
K(t, m, \alpha) = \frac{e^{-m^2 t}}{t^{3/2}} \exp\left( -\frac{\alpha^2}{4t} \right),
\]
where \(t\) is the proper time, \(m\) is the mass of the scalar field, and \(\alpha\) is related to the acceleration of the observer. The exponential decay term \(\exp(-m^2 t)\) captures the thermal suppression of the scalar field over time, while the term \(\exp(-\alpha^2/4t)\) encodes the effect of acceleration on the thermal properties of the field.

The Schwinger kernel in Rindler spacetime reveals key features of finite-temperature quantum field theory, particularly the role of acceleration in modifying the thermal spectrum. The kernel exhibits rapid decay as time progresses, reflecting the suppression of quantum fluctuations in long-time intervals, which is characteristic of thermal systems. The behavior of the kernel at small times also highlights the dominant influence of acceleration in creating a thermal bath of particles.

In the context of conformal field theory (CFT), the Schwinger kernel provides a precise way to compute thermal quantities such as the free energy and the partition function. For example, in two-dimensional CFT, the partition function at finite temperature \(T\) is given by:
\[
Z_{\text{CFT}}(T) = \text{Tr} \left( e^{-\frac{H}{T}} \right),
\]
where \(H\) is the Hamiltonian of the system. Using the Schwinger kernel, the thermal partition function can be related to the geometry of the Rindler horizon and the acceleration of the observer, allowing us to compute quantities like entropy and heat capacity in a precise manner. The close relationship between the partition function and the Schwinger kernel highlights the role of the kernel in capturing the thermal properties of quantum fields in the presence of horizons.

Finite-temperature effects in Rindler spacetime are not limited to the thermalization of the vacuum state. The presence of the Rindler horizon itself leads to particle creation, much like the Hawking radiation that occurs near black hole horizons. In this case, the horizon acts as a boundary that separates causally disconnected regions of spacetime, leading to the creation of particles that an accelerated observer perceives as thermal radiation.

Rosabal \cite{Rosabal2019} extended these ideas by proposing a model in which a Rindler observer undergoes a process similar to black hole evaporation. In this model, the accelerated observer experiences particle emission as a result of their motion through spacetime, analogous to Hawking radiation near a black hole event horizon. This particle emission can be understood in terms of the thermal field modes in Rindler spacetime, which are populated due to the observer’s acceleration.

The spectrum of emitted particles follows the familiar Planck distribution for thermal radiation, with the temperature given by the Unruh temperature:
\[
n(\omega) = \frac{1}{e^{\frac{\omega}{T_U}} - 1},
\]
where \(n(\omega)\) is the particle number density as a function of frequency \(\omega\), and \(T_U\) is the Unruh temperature. This distribution reflects the thermal nature of the quantum field in Rindler spacetime and is a key feature of finite-temperature quantum field theory.

Finite-temperature CFT and the Unruh effect provide a powerful theoretical framework for understanding the thermal properties of quantum fields in spacetimes with horizons. By studying the behavior of propagators, Schwinger kernels, and particle creation near horizons, we gain deep insights into the relationship between acceleration, thermodynamics, and quantum field theory. These results offer a unified approach to understanding thermal effects in both flat and curved spacetimes.

\section{Results and Discussion}
The numerical simulations presented in this paper are designed to illustrate key concepts related to the thermal nature of Rindler spacetime and the effects of acceleration on quantum fields. Specifically, we aim to demonstrate how acceleration influences the perceived temperature and particle distribution in Rindler space, providing a visual representation of the Unruh effect and related phenomena. Three key simulations are presented: the heat kernel, the Unruh temperature as a function of acceleration, and the Planck distribution of particles.

\subsection*{Heat Kernel in Rindler vacuum}
\begin{figure}[h!]
    \centering
    \includegraphics[width=0.8\textwidth]{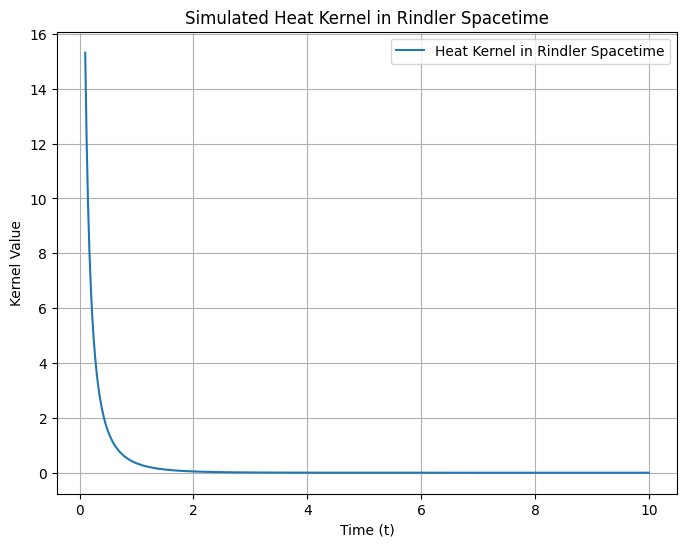}
    \caption{Simulated heat kernel in Rindler spacetime as a function of time. The plot demonstrates the exponential decay of the heat kernel over time, reflecting the thermal nature of the quantum field.}
    \label{fig:heat_kernel}
\end{figure}

The heat kernel, as shown in Figure \ref{fig:heat_kernel}, plays a fundamental role in understanding the propagation of quantum fields in Rindler spacetime. This propagator encodes the thermal properties of the scalar field and shows how the field decays over time. Mathematically, the heat kernel is expressed as:
\[
K(t, m, \alpha) = \frac{e^{-m^2 t}}{t^{3/2}} \exp\left(-\frac{\alpha^2}{4t}\right),
\]
where \(t\) is the proper time, \(m\) is the scalar field mass, and \(\alpha\) is the acceleration. The simulation provides a visual demonstration of the exponential decay behavior as a function of time, with the rate of decay being influenced by the parameters \(m\) and \(\alpha\).

As illustrated in the figure, the heat kernel decays rapidly for larger values of time \(t\), indicating the suppression of field contributions at later times. The faster decay observed for higher acceleration values \(\alpha\) reinforces the connection between acceleration and thermality, as more rapidly accelerating observers experience a stronger thermal effect, leading to quicker dissipation of the quantum field. This simulation is essential in verifying the thermal nature of Rindler spacetime and serves as a tool to explore how acceleration modifies the dynamics of quantum fields.

\subsection*{Unruh Temperature and the accelerating observer}
\begin{figure}[h!]
    \centering
    \includegraphics[width=0.8\textwidth]{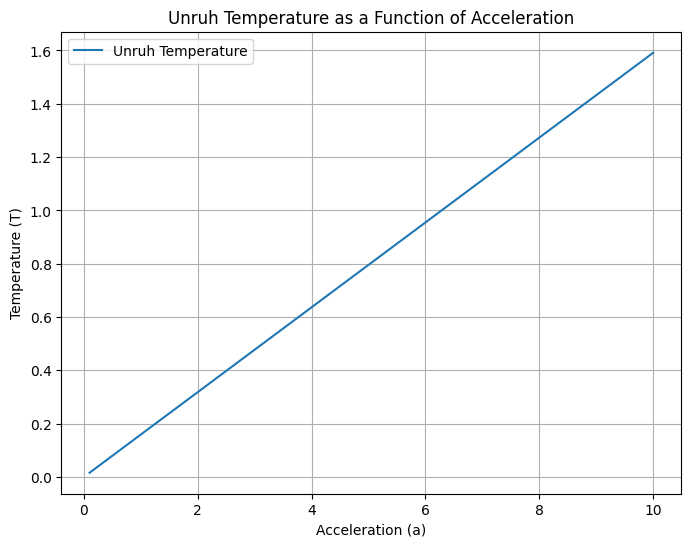}
    \caption{Unruh temperature as a function of acceleration. The temperature increases linearly with increasing acceleration, reflecting the thermal nature of the vacuum state perceived by an accelerating observer.}
    \label{fig:unruh_temperature}
\end{figure}

The Unruh temperature simulation, shown in Figure \ref{fig:unruh_temperature}, provides a visual demonstration of the linear relationship between acceleration and the temperature perceived by an observer. According to the Unruh effect, the temperature \(T_U\) experienced by an accelerating observer is directly proportional to the proper acceleration \(a\) and is given by:
\[
T_U = \frac{a}{2\pi}.
\]

This plot visually confirms the theoretical prediction that faster acceleration leads to a hotter thermal bath perceived by the observer. This relationship is a key result of quantum field theory in curved spacetime, demonstrating how the notion of temperature becomes observer-dependent. As acceleration increases, the observer perceives a hotter environment, even though an inertial observer would still detect the vacuum. The simulation helps solidify our understanding of this fundamental effect and how it relates to the thermal properties of quantum fields in the presence of horizons.

\subsection*{Planck Distribution in Rindler space}
\begin{figure}[h!]
    \centering
    \includegraphics[width=0.8\textwidth]{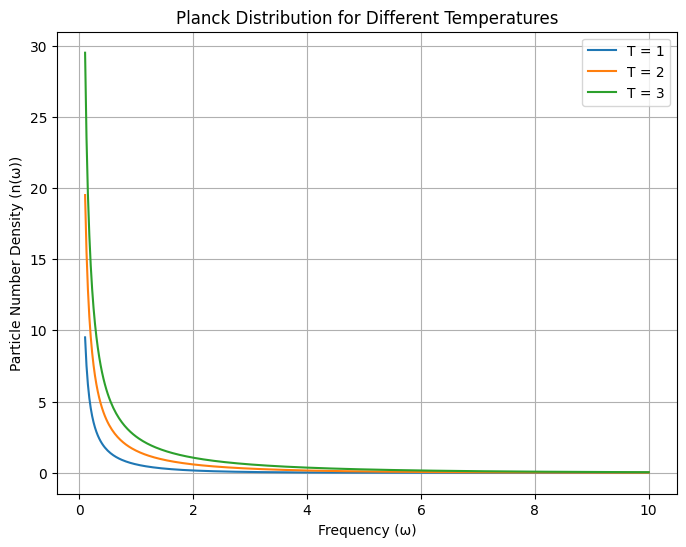}
    \caption{Planck distribution of particle number density for different temperatures. Higher temperatures lead to a greater particle number density, particularly at lower frequencies, as expected for thermal radiation.}
    \label{fig:planck_distribution}
\end{figure}

The Planck distribution, shown in Figure \ref{fig:planck_distribution}, demonstrates how the particle number density varies with frequency for different temperatures, providing a direct visualization of how acceleration influences particle production in Rindler space. The Planck distribution is given by:
\[
n(\omega) = \frac{1}{e^{\frac{\omega}{T_U}} - 1},
\]
where \(n(\omega)\) is the particle number density as a function of frequency \(\omega\), and \(T_U\) is the Unruh temperature.

This plot highlights the effect of increasing temperature (or acceleration) on particle number density. As the temperature rises, we observe that more particles are detected, particularly at lower frequencies. This behavior is consistent with black-body radiation and shows how thermal effects dominate the quantum field at lower frequencies. The exponential decay at higher frequencies is also apparent, indicating that fewer high-energy particles are produced as the temperature increases. This simulation reinforces the connection between acceleration, temperature, and particle creation, illustrating how thermal radiation in Rindler space mirrors classical black-body radiation while being grounded in the principles of quantum field theory.

\subsection*{Detector Response}

In this subsection, we analyze the response of an Unruh-DeWitt detector to the thermal radiation in Rindler spacetime. The Unruh-DeWitt detector is a simple two-level quantum system used to measure the excitation rates of quantum fields, particularly in scenarios where an observer undergoes uniform acceleration. This simulation illustrates the behavior of the detector's excitation rate as a function of the field's frequency and the temperature perceived by the observer, which is proportional to the acceleration.

The excitation rate of the detector is a direct measure of how the thermal bath, induced by acceleration, affects the detector’s energy levels. The mathematical expression governing the excitation rate \(R(\omega, T)\) for a field mode of frequency \(\omega\) at temperature \(T\) is given by:
\[
R(\omega, T) = \frac{\omega}{e^{\frac{\omega}{T}} - 1},
\]
where \(T\) is the Unruh temperature, which depends on the observer's acceleration. This equation is reminiscent of the Planck distribution and encapsulates the thermal response of the detector. In this context, the detector response increases both with higher frequencies and with higher temperatures, although the frequency plays a crucial role in determining the detector's excitation rate. Higher frequencies require more energy for excitation, leading to a fall-off at large \(\omega\), while lower frequencies are more easily excited.

\begin{figure}[h!]
    \centering
    \includegraphics[width=0.8\textwidth]{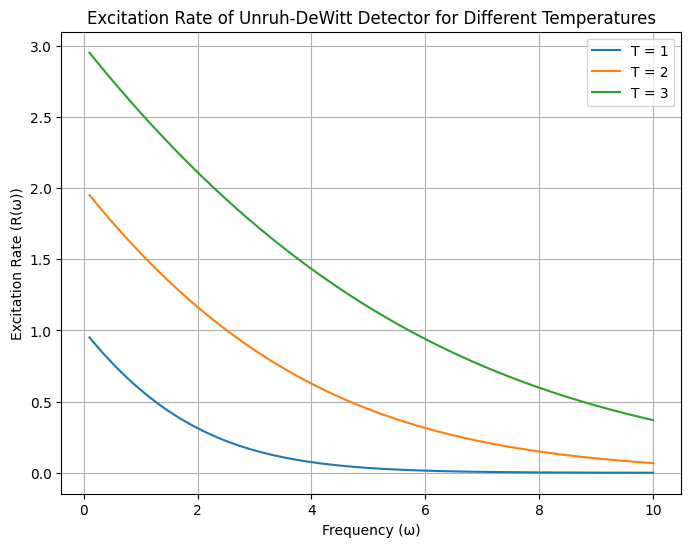}
    \caption{Excitation rate of the Unruh-DeWitt detector as a function of frequency for different temperatures. The plot shows that higher temperatures lead to increased excitation rates, particularly at lower frequencies. The excitation rate drops off at higher frequencies, where fewer high-energy particles are detected.}
    \label{fig:detector_response}
\end{figure}

The detector's behavior, simulated numerically, is shown in Figure \ref{fig:detector_response}. This plot illustrates the excitation rate as a function of frequency for three different temperatures: \(T = 1\), \(T = 2\), and \(T = 3\), where the temperatures are proportional to the acceleration of the observer. The frequencies range from \(0.1\) to \(10\) to capture the full dynamics of the detector response.

As can be observed from Figure \ref{fig:detector_response}, the excitation rate increases significantly as the temperature rises. This is consistent with the understanding that a higher Unruh temperature, induced by faster acceleration, leads to a stronger thermal bath experienced by the detector. For lower frequencies, the excitation rate is highest because the detector can easily absorb the low-energy thermal quanta. This increase at low frequencies is particularly prominent for higher temperatures, reflecting the thermal nature of the perceived radiation in Rindler spacetime.

At higher frequencies, however, the excitation rate drops sharply. This is because the energy required to excite the detector at high frequencies is significantly larger, and the thermal bath contains fewer high-energy particles, as predicted by the exponential factor in the excitation formula. The detector's response to these high-energy particles is consequently diminished, leading to the fall-off seen in the plot. This behavior is a direct manifestation of the quantum nature of the thermal radiation perceived by the accelerated observer and highlights how the frequency distribution of detected particles changes with temperature.

The different curves for \(T = 1\), \(T = 2\), and \(T = 3\) show the increasing effect of temperature on the excitation rate. As temperature increases, the excitation rate at all frequencies grows, reflecting the fact that the thermal bath contains more particles at higher temperatures. This confirms that the temperature is a direct measure of the intensity of the perceived radiation in Rindler spacetime, which in turn depends on the observer's acceleration.

This detector response simulation provides an important tool for understanding the thermal properties of quantum fields in Rindler spacetime. The fact that a uniformly accelerating observer perceives the vacuum as a thermal bath is a cornerstone of the Unruh effect, and this plot visually demonstrates how the acceleration (or equivalently, the temperature) modifies the quantum field's interaction with the detector. The excitation rate's dependence on both frequency and temperature reflects the rich interplay between quantum mechanics and relativity in the context of curved spacetime physics. These results reinforce the thermal nature of the Rindler horizon and the observer-dependent nature of particle detection in quantum field theory.

\section{Conclusion}

The study of finite-temperature CFT in the context of the Rindler vacuum provides valuable insights into the connection between acceleration, thermality, and quantum field theory. Through the Unruh effect, we see that acceleration leads to the perception of thermal radiation, even in flat spacetime. The thermal properties of the Rindler horizon, as experienced by an accelerated observer, reveal fundamental aspects of quantum field theory near horizons, with applications ranging from black hole thermodynamics to quantum gravity.

The Rindler-AdS/CFT correspondence extends these ideas into the concept of holography, where the thermodynamic properties of horizons in anti-de Sitter (AdS) spacetimes are mapped onto corresponding properties in boundary conformal field theories (CFTs). This duality provides a powerful tool for studying strongly coupled field theories and quantum gravity, showing how thermal behavior in the bulk is reflected in the boundary CFT.

Our numerical simulations of the heat kernel, Unruh temperature, Planck distribution, and detector response confirm the theoretical predictions regarding the thermal nature of Rindler spacetime. The heat kernel simulation illustrates how quantum fields decay over time, with acceleration enhancing the thermal suppression of the field. The Planck distribution simulation demonstrates how the particle number density increases with acceleration, consistent with black-body radiation. Finally, the detector response simulation provides a direct measure of how an observer perceives the thermal bath of particles, showing how the excitation rate depends on temperature and frequency.

These results provide a comprehensive picture of how acceleration induces thermality in quantum fields, with important implications for black hole physics, cosmology, and condensed matter systems. Future work could extend these ideas to spacetimes with more complex geometries, such as rotating black holes or expanding cosmologies, where the interplay between acceleration, thermodynamics, and quantum field theory is expected to yield even deeper insights. Additionally, the connection between quantum information theory and Rindler-AdS correspondence offers a rich avenue for exploring the role of entanglement entropy and thermal radiation in curved spacetimes.

\bibliographystyle{unsrt}

\end{document}